\documentclass[a4paper,11pt]{article}
\usepackage{pos}
\usepackage{physics}
\usepackage[section]{placeins}
\usepackage{lineno}
\let\OLDthebibliography\thebibliography
\renewcommand\thebibliography[1]{
  \OLDthebibliography{#1}
  \setlength{\parskip}{0pt}
  \setlength{\itemsep}{0pt}
}
\title{Inclusive Jet Measurements in Pb--Pb Collisions at 5.02~TeV with ALICE using Machine Learning Techniques}

\author*[a]{Hannah Bossi for the ALICE Collaboration}

\affiliation[a]{Yale University,\\
  Wright Laboratory, New Haven, CT, USA}

\emailAdd{hannah.bossi@cern.ch}

\abstract{These proceedings report on measurements of the jet spectrum and nuclear modification factor for inclusive full jets (containing both charged and neutral constituents) in Pb--Pb and \textit{pp} collisions at $\sqrt{s_{\rm NN}} = 5.02$ TeV recorded with the ALICE detector. These measurements use a machine learning based background correction \cite{methodPaper}, which reduces residual fluctuations. This method allows for measurements to lower transverse momenta and larger jet resolution parameter (\textit{R}) than previously possible in ALICE. In this method, machine learning techniques are used to correct the jet transverse momentum on a jet-by-jet basis using jet parameters such as information about the constituents of the jet. Studies that investigate the effect of the potential fragmentation bias introduced by learning from constituents will also be discussed. With these studies in mind, the results are compared to theoretical predictions. \\
}

\FullConference{%
  HardProbes2020\\
  1-6 June 2020\\
  Austin, Texas}

\begin{document}
\maketitle

\section{Introduction}
Measurements of inclusive jet suppression (or $R_{\rm AA}$ as defined in Equation \ref{Eq:Raa} as the ratio of the per-event jet yield in Pb–Pb and the cross section in \textit{pp} multiplied by $T_{\rm AA}$ accounting for the collision geometry) serve to search for signatures of jet quenching in heavy-ion (HI) collisions,
\begin{equation}\label{Eq:Raa}
    R_{AA} = \frac{\frac{1}{\expval{T_{\rm AA}}}\frac{1}{\it{N}_{\rm events, PbPb}}\frac{\rm d^{2}\it{N}_{\rm jet}^{\rm AA}}{\rm d\it{p}_{\rm T} \rm d\eta}}{\frac{\rm d^{2}\sigma_{\rm pp}}{\rm d\it{p}_{\rm T} \rm d\eta}}.
\end{equation}

The nature and extent of this energy loss is expected to vary for different jet resolution parameters $R$ across $p_{\rm T}$ scales. Jet quenching models describe the $R$-dependence of the $R_{\rm AA}$ differently, with some models predicting this dependence to be stronger at lower jet $p_{\rm T}$ \cite{hybridModel,SCETg, LBT, JEWEL}. Recent experimental efforts have extended measurements at high $p_{\rm T}$ to $R = 1.0$  \cite{CMS_LargeR}, but regions of large $R$ and low $p_{\rm T}$ are also of interest.

Reconstructing the jet transverse momentum ($p_{\rm T}$) in HI collisions is challenging due to the large fluctuating background from the underlying event (UE). For example, the fluctuations in the charged particle momentum density in central (0--10$\%$) Pb--Pb collisions at the LHC are $\approx$ 18 GeV/$c$ \cite{AB}. Lower $p_{\rm T}$ UE jets, commonly called fake jets, additionally contaminate the signal at low $p_{\rm T}$. One common treatment of this background (herein referred to as the area-based or AB method) is to subtract off the event-averaged momentum density (excluding the two leading jets) multiplied by the jet area from the uncorrected jet $p_{\rm T}$ \cite{AB}.

A leading track cut of 7 GeV/$c$ for $R$ = 0.4 jets is typically applied in order to remove the fake jet contribution at the cost of an introduced bias. While the AB method effectively corrects for the average background, it does not account for region-to-region fluctuations. These residual fluctuations are commonly handled in an unfolding procedure. Due to the limitations of the large background, lower jet $p_{\rm T}$ and large $R$ are less studied regions with inclusive jet probes.
 
In these proceedings, inclusive jet measurements with ALICE utilizing a machine learning (ML) based method previously developed for charged particle jets \cite{methodPaper} will be discussed. This ML-based method allowed for the extension of charged particle jet measurements to lower $p_{\rm T}$ for $R$ = 0.6 jets, unprecedented in ALICE. These proceedings will present the extension of this method to full jets, which is desirable as full jets are more aligned with the traditional definition of a jet.

\section{Analysis Details}
The results discussed in these proceedings utilize Pb--Pb data collected with the ALICE detector \cite{alice} in 2015 at $\sqrt{s_{\rm NN}} = 5.02$ TeV with an integrated luminosity of 0.4 $n \text{b}^{-1}$. Central events (0--10\%) were analysed with a minimum bias trigger setup. The training of the ML estimator (see Section \ref{Sec_ML}) and the response matrix used for unfolding utilize \textit{detector level} events and \textit{hybrid level} events. These hybrid events were created using a PYTHIA8 \cite{pythia} generated events propagated through a GEANT3 \cite{geant} simulation of the ALICE detector (herein referred to as the \textit{detector level}) which were embedded into real Pb--Pb data to mimic HI background effects (herein referred to as the \textit{hybrid level}). The ML approach will be compared with results from \cite{jamesAB} which are based on the AB method. For the evaluation of $R_{\rm AA}$, the \textit{pp} reference spectrum without a leading track bias was taken from \cite{jamesAB}.

These proceedings focus on full jets that contain both the charged and neutral component of the jet measured using charged particles registered by the Inner Tracking System \cite{its} and the Time Projection Chamber \cite{tpc} and electromagnetic clusters measured from the Electromagnetic Calorimeter \cite{emcal}. The FastJet package \cite{fastjet} was used to cluster charged constituents above 150 MeV/$c$ and neutral constituents above 300 MeV/$c$ into jets using the anti-$k_{\rm T}$ algorithm \cite{antikt} with a $p_{\rm T}$ recombination scheme. 

\section{ML-estimator}\label{Sec_ML}
The ML-based background estimator creates a mapping between the measured and reconstructed jet $p_{\rm T}$, correcting each jet for the background particles that overlay it. The ML estimator itself employs a shallow neural network as implemented in scikit-learn \cite{scikit} with three layers. As input the estimator utilized a minimal, discriminative set of features including jet properties and the properties of the constituents of the jet. 

\begin{figure}[ht!]
    \centering
    \includegraphics[scale = 0.16]{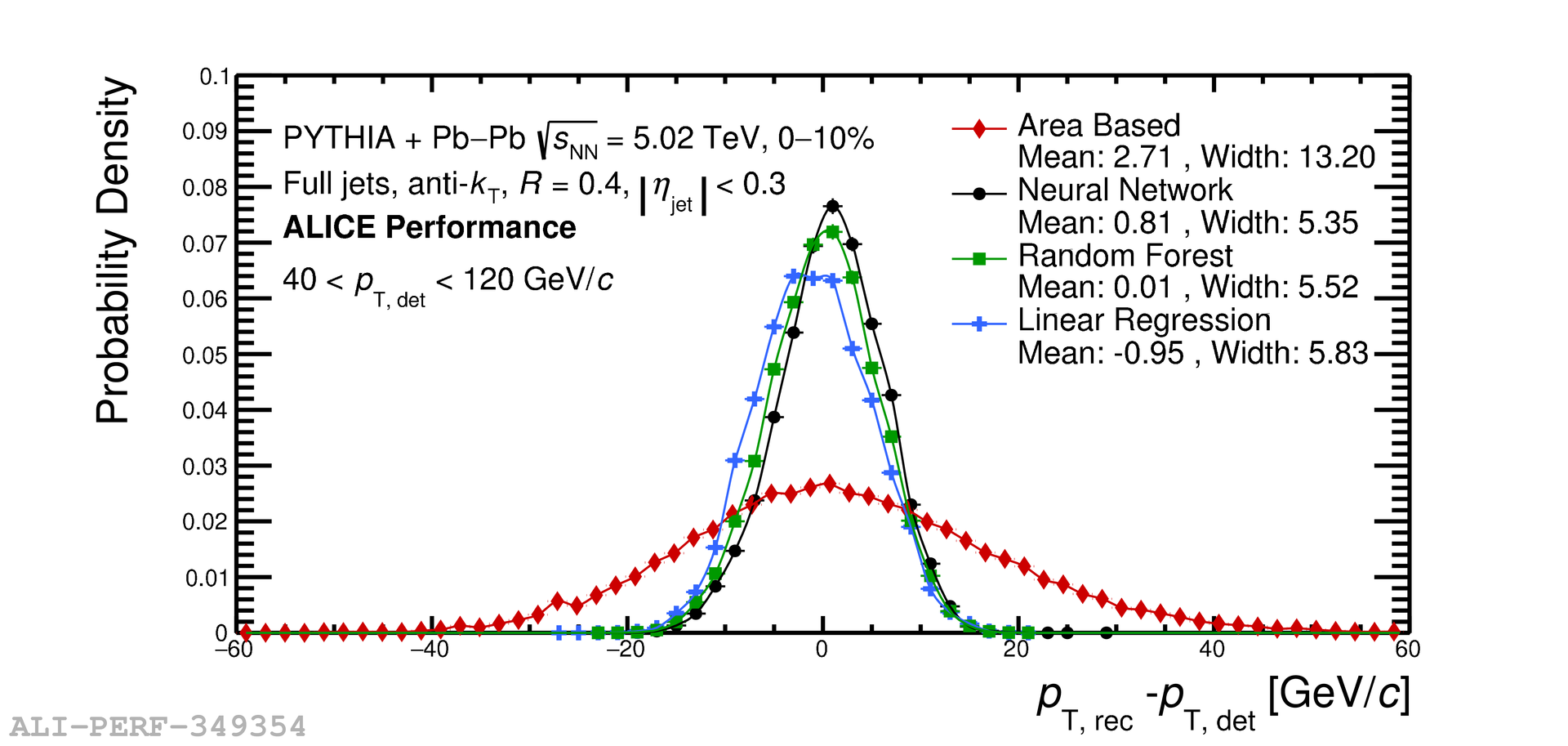}
    \caption{The $\delta{p_{\rm T}} = p_{\rm T, rec} - p_{\rm T, det}$ distributions for $R = 0.4$ central (0--10$\%$) jets.}
    \label{fig:residuals}
\end{figure}

The ML estimator is a regression task where the regression target is the reconstruction of the detector level PYTHIA jet $p_{\rm T}$ from the hybrid event used in training the estimator. The performance of the estimator is quantified with $\delta p_{\rm T}$ distributions measuring the difference between the jet $p_{\rm T}$ predicted by the ML estimator and the PYTHIA detector level jet $p_{\rm T}$. A narrowing of the width in $\delta p_{\rm T}$ corresponds to reduction of residual fluctuations after background subtraction. Figure \ref{fig:residuals} demonstrates that the ML estimators more effectively reduce the residual fluctuations remaining after background subtraction than the AB method and leads to a reduced width of the $\delta{p_{\rm T}}$ distribution.

\section{Results}
\begin{figure}[ht]
    \centering
    \includegraphics[width = 0.7\textwidth]{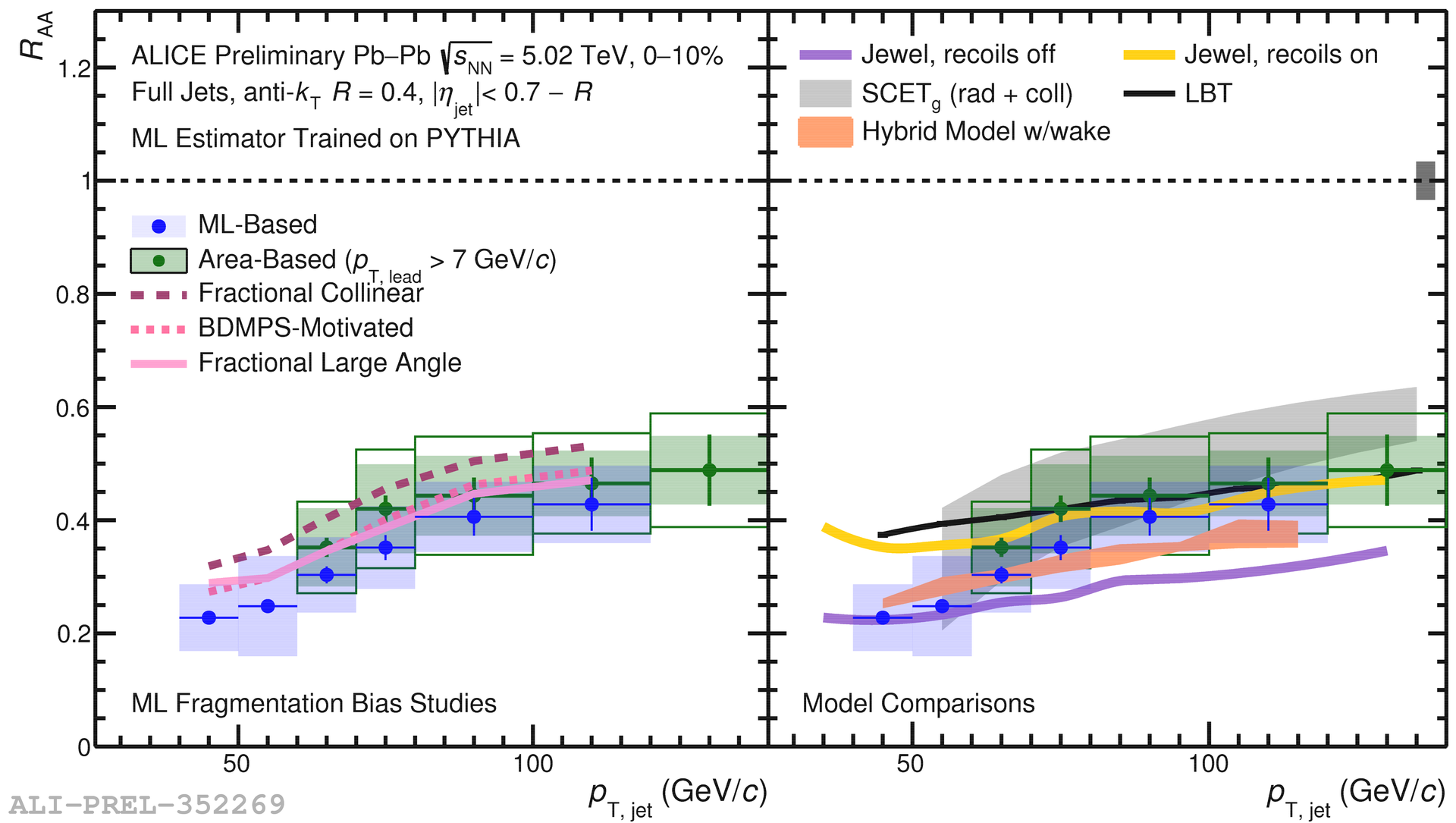}
    \caption{The $R_{\rm AA}$ with the ML and AB method are show for $R = 0.4$ jets in 0--10$\%$ collisions. In the left panel, the $R_{\rm AA}$ are compared to the fragmentation bias curves for a toy model with three different modifications. The right panel compares the $R_{\rm AA}$ to various theoretical predictions. The ML-based $R_{\rm AA}$ and the AB $R_{\rm AA}$ \cite{jamesAB} are compared to the same \textit{pp} reference, which does not include a leading track bias.}
    \label{fig:results}
\end{figure}

The ML method was applied to full $R= 0.4$ jets in central (0--10$\%$) Pb--Pb collisions in order to measure the inclusive jet spectrum. The $R_{\rm AA}$  was then evaluated using this inclusive jet spectrum and the scaled \textit{pp} reference spectrum from \cite{jamesAB}.
These results are shown in Figure \ref{fig:results} compared to the previous result obtained using the AB method \cite{jamesAB}. With the ML-based correction, measurements of jet suppression are extended to 40 GeV/\textit{c}. The systematic uncertainties were also reduced.

The left panel of Figure \ref{fig:results} illustrates an investigation of the fragmentation bias of the ML-based estimator. This bias is introduced by the inclusion of constituent information in training. Training on PYTHIA jets causes a bias towards a \textit{pp} PYTHIA fragmentation which is suggested to differ from the fragmentation pattern of jets in Pb--Pb by numerous studies (e.g.\cite{cmsFF, atlasFF}).  For these investigations, a toy model is used with three different modifications to the constituents of the jet and therefore the fragmentation function. The fractional methods refer to \textbf{each} constituent losing a fraction of its energy at specified angles from the jet axis. The BDMPS method is a modification where the emission angle and energy is selected by sampling the BDMPS gluon emission spectrum \cite{BDMPS:A, BDMPS:B,BDMPS}. The potential fragmentation bias of such modifications is quantified by training the ML on the modified toy model and calculating the resulting $R_{\rm AA}$. Figure \ref{fig:results} demonstrates the relative robustness of this method to the explored biases.

Keeping these studies in mind, comparisons with theoretical predictions are shown in the right panel of Figure \ref{fig:results}. Comparisons with JEWEL with recoils on/off \cite{JEWEL}, SCETg (with radiative and collisional energy loss) \cite{SCETg}, Hybrid model (with wake) \cite{hybridModel}, and LBT \cite{LBT} are shown. This measurement of inclusive jets to lower jet $p_{\rm T}$ constrains models in a less studied region.

\FloatBarrier
\section{Conclusion}
In these proceedings, a ML based background correction is presented, which allows for the extension of inclusive jet suppression measurements down to 40 GeV/$c$. The fragmentation bias introduced by training the ML on the constituents from PYTHIA has been quantified, which also demonstrates the robustness of this method. Future analyses with this method will explore larger resolution parameters and extend tests of the fragmentation bias.


\begin{thebibliography}{23}
\bibitem{methodPaper}
R. Haake, C. Loizides, "Machine-learning-based jet momentum reconstruction in heavy-ion collisions," \textit{Phys. Rev. C} 99, 064904 (2019). 
\bibitem{hybridModel}
D. Pablos, "Jet Suppression From a Small to Intermediate to Large Radius," \textit{Phys. Rev. Lett.} 124, 052301 (2020).
\bibitem{JEWEL}
R. K. Elayavallli, K. C. Zapp, "Medium response in JEWEL and its impact on jet shape observables in heavy ion collisions" \textit{JHEP} 1707 (2017) 141.

\bibitem{SCETg}
H. T. Li, I. Vitev, "Inclusive heavy flavor jet production with semi-inclusive jet functions: from proton to heavy-ion collisions" \textit{JHEP} 1007 (2019) 148.

\bibitem{LBT}
He et. al. "Interplaying mechanisms behind single inclusive jet suppression in heavy-ion collisions, "\textit{Phys. Rev. C.} 99, 054911 (2019).

\bibitem{CMS_LargeR}
CMS Collaboration, "Measurement of Jet Nuclear Modification Factor in Pb--Pb Collisions at  $\sqrt{s_{\rm NN}}$ = 5.02 TeV with CMS," CMS-PAS-HIN-18-014, \texttt{http://cds.cern.ch/record/2698506}.

\bibitem{AB}
ALICE Collaboration, "Measurement of Event Background Fluctuations For Charged Particle Jet Reconstruction in Pb--Pb Collisions at $\sqrt{s_{\rm NN}}$ = 2.76 TeV," \textit{JHEP} 1203 (2012) 053.

 

\bibitem{alice}
ALICE Collaboration, "The ALICE experiment at the CERN LHC", J. Instrum. 3 (2008) 

\bibitem{pythia}
T. Sjöstrand, S. Mrenna and P. Skands, /textit{Comput. Phys. Comm.} 178 (2008) 852.

\bibitem{geant}
R. Brun et al.: GEANT Detector Description and Simulation Tool, CERN Program Library Long
Writeup CERN-W-5013 (1994).

\bibitem{jamesAB}
ALICE Collaboration, "Measurements of inclusive jet spectra in pp and central Pb--Pb collisions at $\sqrt{s_{\rm NN}} = 5.02$", \textit{Phys. Rev. C.} 101, 034911 (2020)

\bibitem{its}
ALICE Collaboration, "Alignment of the ALICE Inner Tracking System with cosmic ray tracks," \textit{JISNT}, 5:P03003, 2010

\bibitem{tpc}
J. Alme, et al., "The ALICE TPC, a large 3-dimensional tracking device with fast readout for ultra-high multiplicity events," \textit{Nucl.Instrum.Meth.}, A622:316–367, 2010.

\bibitem{emcal}
ALICE Collaboration, "ALICE electromagnetic calorimeter technical design report," CERN-ALICE-TDR-014, CERN-LHCC-2008-014, 2008 \href{https://inspirehep.net/literature/794183}{[https://inspirehep.net/literature/794183]}

\bibitem{fastjet}
M. Cacciari, G.P. Salam and G. Soyez, \textit{Eur.Phys.J.} C72 (2012) 1896.

\bibitem{antikt}
M. Cacciari, G.P. Salam, and G. Soyez, "The anti-kT jet clustering algorithm", \textit{JHEP} 0804 (2008) 063.

\bibitem{scikit}
Pedregosa et al, "Scikit-learn: Machine Learning in Python," JMLR 12, pp. 2825--2830 (2011).

\bibitem{cmsFF}
CMS Collaboration, "Measurement of jet fragmentation in PbPb and pp collisions at $\sqrt{s_{\rm NN}}$ = 2.76 TeV," \textit{Phys. Rev. C.} 90, 024908 (2014).

\bibitem{atlasFF}
ATLAS Collaboration, "Measurement of jet fragmentation in Pb+Pb and pp collisions at $\sqrt{s_{\rm NN}}$ = 5.02 TeV with the ATLAS Detector," \textit{Phys. Rev. C.} 98, 024908 (2018).

\bibitem{BDMPS:A}
R. Baier, Yu. L. Dokshitzer, A. H. Mueller, S. Peigne,  D. Schiff, "Radiative energy loss of high-energy quarks and gluons in a finite volume quark-gluon plasma", \textit{Nucl. Phys.} B483 (1997) 291--320.

\bibitem{BDMPS:B}
R. Baier, Yu. L. Dokshitzer, A. H. Mueller, S. Peigne,  D. Schiff,"Radiative energy loss and p(T) broadening of high-energy partons in nuclei," \textit{Nucl. Phys.} B484 (1997) 265--282.

\bibitem{BDMPS}
R. Baier, Yu. L. Dokshitzer, A. H. Mueller, D. Schiff, "Quenching of hadron spectra in media," \textit{JHEP} 0109 (2001) 033



\end{thebibliography}
\end{document}